# Propagation properties of partially coherent electromagnetic hyperbolic-sine-Gaussian vortex beam through anisotropic atmospheric turbulence

JIN CAO[1,2,3], RUFENG TANG[1], KAI HUANG[2], YUQIANG LI[1,*] AND YONGGEN XU[4]

[1] *Yunnan Observatories, Chinese Academy of Sciences, Kunming, Yunnan 650216, China*
[2] *College of Mathematics and Physics, Leshan Normal University, Leshan, Sichuan 614000, China*
[3] *University of Chinese Academy of Sciences, Beijing 100049, China*
[4] *Shool of Science, Xihua University, Chengdu, Sichuan 610039, China*
*\* Corresponding author: lyq@ynao.ac.cn*

**Abstract:** Utilizing the extended Huygens-Fresnel principle and the Rytov approximation, the analytical formula for the propagation of a partially coherent electromagnetic hyperbolic-sine-Gaussian vortex beam (PCEShVB) in anisotropic atmospheric turbulence has been theoretically derived. Detailed studies have been conducted on the evolution characteristics of average intensity, the degree of coherence (DOC) and the degree of polarization (DOP) of the beam in turbulence. The results show that during propagation, the intensity distribution of the beam will exhibit a spiral structure, and the overall distribution of light spots will rotate in a direction related to the sign of the topological charge. The DOC distribution of PCEShVB will display a pattern reminiscent of beam interference fringes with an increase in propagation distance, with the number of 'interference fringes' greatly impacted by the hyperbolic sine parameter. Furthermore, PCEShVB with large initial coherent length and hyperbolic sine parameter will increase the degree of separation of the spots and yield a large DOP. Finally, for the validation of the theoretical findings, the random phase screen method was employed to simulate the propagation of PCEShVB through anisotropic atmospheric turbulence. The studies revealed a consistent alignment between the simulation results and the theoretical predictions.

## 1. Introduction

The laser has found extensive utility in atmospheric applications or free space optical communication, laser remote sensing, and laser radar, owing to its intrinsic attributes of high intensity, good coherence, and excellent monochromaticity [1,2]. However, when deployed as a communication medium in these scenarios, the Earth's atmosphere introduces random turbulence that significantly impacts the laser's propagation, resulting in disturbances such as beam jitter, light intensity flicker, and beam expansion ,These turbulence effects pose a substantial challenge to the effective application of lasers in these contexts. [3–5]. Numerous studies have demonstrated that partially coherent (PC) beams, characterized by low coherence, uniform intensity distribution, and robust propagation stability, are effective at mitigating the impact of atmospheric turbulence [4,6,7]. On the other hand, vortex beams, where each photon carries Orbital Angular Momentum (OAM), offer heightened confidentiality and concealment through OAM data encoding; any tampering with the link results in broadening of the beam's OAM spectrum, thereby rendering the coding inoperable [8,9]. Consequently, vortex beams are better suited for laser communication and other applications. Furthermore, when compared to non-vortex beams, vortex beams more effectively counteract the effects of light intensity fluctuation and spot expansion caused by atmospheric turbulence. Extensive research has also delved into the behavior of various types of vortex beams in atmospheric turbulence [10]. In 1997, Casperson et al. demonstrated that the Hermite-sine-Gaussian beam provides a new solution to the wave equation within the paraxial approximation. They showed that the hyperbolic sinusoidal beam (ShB) theoretically represents a broad category of beams, including the sine (cosine) Gaussian beam and the hyperbolic sine (cosine) Gaussian beam [11].

Subsequently, Huang et al. conducted a comprehensive exploration of the evolution behavior of coherent vortices and average intensity, along with the modifications of the spectral degree of polarization for partially coherent electromagnetic hyperbolic-sine-Gaussian vortex beams propagating through non-Kolmogorov turbulence. Their study sheds light on the impact of non-Kolmogorov turbulence on the propagation and evolution behavior of electromagnetic vortex beams [12]. Additionally, Eyyuboğlu et al. investigated the dynamic evolution of Hermite-sinh-Gaussian laser beams in a turbulent atmosphere [13]. Furthermore, Zhang et al. examined the mean aperture scintillation, mean signal noise, and mean bit error rate of the sinusoidal Gaussian beam during transmission in atmospheric turbulence. They also conducted a comparative study of the sinusoidal Gaussian beam and the sinusoidal Gaussian vortex beam under the same conditions [14]. Following the aforementioned studies, Mert Bayraktar carried out an extensive investigation into the propagation characteristics of the astigmatic hyperbolic sinusoidal Gaussian beam in oceanic turbulence, aiming to provide insights applicable to underwater communication, imaging, and range measurement applications [15]. Subsequently, he also investigated the transmission characteristics of this beam in biological tissues, and the research findings will be utilized in the design of optical medical devices [16]. In summary, the partially coherent electromagnetic hyperbolic-sine-Gaussian vortex beam (PCEShVB) possesses profound scientific research value and considerable potential for applications in various fields, thus this paper will conduct a more in-depth study of this optical beam.

For an extended period, the investigation of atmospheric turbulence has primarily relied on the Kolmogorov spectrum model, assuming uniform and isotropic turbulence. The Kolmogorov turbulence spectrum is generally correct in the atmospheric boundary layer (approximately 1 to 2km above the Earth's surface), where it exhibits uniformity and isotropy. However, recent advances in turbulence observation experiments have revealed strong evidence that optical turbulence in the free atmosphere displays significant heterogeneity and anisotropy [17]. The characteristics of atmospheric turbulence in this region deviate from the classical Kolmogorov model. Consequently, Toselli et al. pioneered a non-Kolmogorov turbulence spectrum model that accounts for the impacts of inner and outer turbulence scales. Furthermore, anisotropy factors were introduced to capture the anisotropic nature of atmospheric turbulence, thereby enhancing the comprehensive understanding of atmospheric turbulence theory [18,19]. To date, there is a lack of reported research on the propagation properties of PCEShVB through anisotropic atmospheric turbulence.

In this study, the extended Huygens-Fresnel principle was employed to derive the cross-spectral density matrix of PCEShVB propagating in anisotropic atmospheric turbulence. A comprehensive analysis of the propagation characteristics, encompassing average intensity, degree of coherence (DOC), and degree of polarization (DOP) of the beam in atmospheric turbulence, was conducted. Finally, for the validation of the theoretical findings, the random phase screen method was employed to simulate the propagation of PCEShVB through anisotropic atmospheric turbulence. The investigations revealed a consistent alignment between the simulation outcomes and the theoretical predictions.

## 2. Theoretical model

The electric field of a full coherent hyperbolic-sine-Gaussian vortex beam (ShVB) in cartesian coordinates is expressed as [13,20]:

$$E(\boldsymbol{\rho}; z=0) = \exp\left(-\frac{\boldsymbol{\rho}^2}{w_0}\right) \sinh\left[\Omega_0(\rho_x + \rho_y)\right]\left[\rho_x \pm i\,\text{sgn}(m)\rho_y\right]^{|m|}, \quad (1)$$

where $\boldsymbol{\rho} = (\rho_x, \rho_y)$ is the arbitrary transverse position vector in initial plane, $w_0$ is the beam waist for the fundamental Gaussian beam, $m$ denotes the topological charge, $\sinh(\cdot)$ and $\text{sgn}(\cdot)$ represent the hyperbolic sine function and sign function respectively, and $\Omega_0$ is the parameter related to the sinh part of the beam with the units of m$^{-1}$. And '±' takes '−' and '+' when $m$ is positive and negative, respectively.

For partially coherent vector beam, its statistical properties can be represented by a 2×2 cross spectral density matrix (CSDM) as follow [21]:

$$\overleftrightarrow{W}^{(0)}(\boldsymbol{\rho}_1,\boldsymbol{\rho}_2;0) = \begin{bmatrix} W_{xx}(\boldsymbol{\rho}_1,\boldsymbol{\rho}_2;0) & W_{xy}(\boldsymbol{\rho}_1,\boldsymbol{\rho}_2;0) \\ W_{yx}(\boldsymbol{\rho}_1,\boldsymbol{\rho}_2;0) & W_{yy}(\boldsymbol{\rho}_1,\boldsymbol{\rho}_2;0) \end{bmatrix}, \quad (2)$$

Where

$$W_{\alpha\beta}(\boldsymbol{\rho}_1,\boldsymbol{\rho}_2;0) = \langle E_\alpha^*(\boldsymbol{\rho}_1;0) E_\beta(\boldsymbol{\rho}_2;0) \rangle_m, (\alpha,\beta = x,y). \quad (3)$$

In Eq. (2), $\boldsymbol{\rho}_1 = (\rho_{x1}, \rho_{y1})$ and $\boldsymbol{\rho}_2 = (\rho_{x2}, \rho_{y2})$ denote two arbitrary transverse position vectors in source plane. And in Eq. (3), $E_x$ and $E_y$ are the two components of the stochastic electric vector in $x$ and $y$ directions, $\langle \cdot \rangle_m$ denotes the ensemble average, "*" represents the complex conjugate.

Subsequently, an analysis is conducted on electromagnetic hyperbolic-sine-Gaussian vortex beams, further extending the examination to encompass the partially coherent form, with the coherent structure adhering to a Gaussian distribution. So, the elements of the CSDM for PCEShVB in initial plan can be written as:

$$W_{\alpha\beta}(\boldsymbol{\rho}_1,\boldsymbol{\rho}_2;0) = B_{\alpha\beta} \left[ (\rho_{1x}\rho_{2x} + \rho_{1y}\rho_{2y}) \pm i\,\text{sgn}(m)(\rho_{1x}\rho_{2y} - \rho_{2x}\rho_{1y}) \right]^{|m|} \exp\left(-\frac{\boldsymbol{\rho}_1^2 + \boldsymbol{\rho}_2^2}{w_0^2}\right) \\ \times \sinh\left[\Omega_0(\rho_{1x}+\rho_{1y})\right] \sinh\left[\Omega_0(\rho_{2x}+\rho_{2y})\right] \exp\left[-\frac{(\boldsymbol{\rho}_1-\boldsymbol{\rho}_2)^2}{2\sigma_{\alpha\beta}^2}\right], \quad (4)$$

and here, $\delta_{\alpha\beta}$ represents the initial auto-correlation length (when $\alpha=\beta$) and the initial mutual correlation length (when $\alpha\neq\beta$), $B_{\alpha\beta}$ is the complex correlation coefficient used for describing the correlation degree between two electric field components in source plane. In addition, for a partially coherent beam with Gaussian coherent structure, the initial correlation length $\delta_{\alpha\beta}$ must be satisfied with:

$$\max\{\sigma_{xx},\sigma_{yy}\} \leq \sigma_{xy} \leq \min\left\{\frac{\sigma_{xx}}{\sqrt{B_{xy}}}, \frac{\sigma_{yy}}{\sqrt{B_{xy}}}\right\}. \quad (5)$$

Assuming propagation of the beam over a distance $z$ in atmospheric turbulence, the analysis is conducted based on the extended Huygens-Fresnel principle [22,23], the elements of CSDM for PCEShVB on the received plane can be expressed as follows:

$$W_{\alpha\beta}(\boldsymbol{s}_1,\boldsymbol{s}_2;z) = \left(\frac{k}{2\pi z}\right)^2 \iint\iint W_{\alpha\beta}(\boldsymbol{\rho}_1,\boldsymbol{\rho}_2,0) \exp\left\{-\frac{ik}{2z}\left[(\boldsymbol{s}_1-\boldsymbol{\rho}_1)^2 - (\boldsymbol{s}_2-\boldsymbol{\rho}_2)^2\right]\right\} \\ \times \left\langle \exp\left[\psi^*(\boldsymbol{s}_1,\boldsymbol{\rho}_1) + \psi(\boldsymbol{s}_2,\boldsymbol{\rho}_2)\right]\right\rangle_m d^2\boldsymbol{\rho}_1 d^2\boldsymbol{\rho}_2, \quad (6)$$

where, $\boldsymbol{s}_1 = (s_{1x}, s_{1y})$ and $\boldsymbol{s}_2 = (s_{2x}, s_{2y})$ are positions of two points in the received plane, $k=2\pi/\lambda$ is the wave number with $\lambda$ being the wavelength. The term '$\langle \exp[\psi^*(\boldsymbol{s}_1,\boldsymbol{\rho}_1) + \psi(\boldsymbol{s}_2,\boldsymbol{\rho}_2)]\rangle_m$' represents the second-order statistical average of complex phase disturbances caused by atmospheric turbulence, according to the Kolmogorov statistical theory of turbulence and the quadratic approximation of the Rytov phase structure function [24,25], it can be given by:

$$\left\langle \exp\left[\psi^*(\boldsymbol{s}_1,\boldsymbol{\rho}_1) + \psi(\boldsymbol{s}_2,\boldsymbol{\rho}_2)\right]\right\rangle_m \\ = \exp\left\{-4\pi^2 k^2 z \int_0^1 d\xi \int_0^\infty \left\{1 - J_0\left[\kappa|\xi(\boldsymbol{\rho}_1-\boldsymbol{\rho}_2) + (1-\xi)(\boldsymbol{s}_1-\boldsymbol{s}_2)|\right]\right\} \Phi_n(\kappa)\kappa^3 d\kappa \right\}, \quad (7)$$

where $J_0(\cdot)$ is the zero-order Bessel function of the first kind, $\xi = 1-z/L$ is the normalized distance parameter, with $L$ being the total propagation distance, $\Phi_n(\kappa)$ represents the spatial power spectrum of the refractive-index fluctuations of atmospheric turbulence with $\kappa$ denoting the spatial wavenumber. Generally, a first order approximation to $J_0(\cdot)$ is employed, thereby simplifying Eq. (7) as follows:

$$\left\langle \exp\left[\psi^*(s_1, \rho_1) + \psi(s_2, \rho_2)\right] \right\rangle_m$$
$$\cong \exp\left\{ -\frac{\pi^2 k^2 z}{3}\left[(s_1 - s_2)^2 + (s_1 - s_2)(\rho_1 - \rho_2) + (\rho_1 - \rho_2)^2\right]\int_0^\infty \kappa^3 \Phi_n(\kappa)\kappa d\kappa \right\}, \quad (8)$$

Then in this paper, the anisotropic non-Kolmogorov power spectrum is used as a model for atmospheric turbulence, and its spatial power spectrum $\Phi_n(\kappa)$ is given by [26]:

$$\Phi_n(\kappa_x, \kappa_y, \kappa_z) = \frac{\xi_x \xi_y \tilde{C}_n^2 A(\mu)}{\left(\xi_x^2 \kappa_x^2 + \xi_y^2 \kappa_y^2 + \kappa_z^2 + \kappa_0^2\right)^{\frac{\mu}{2}}} \exp\left(-\frac{\xi_x^2 \kappa_x^2 + \xi_y^2 \kappa_y^2 + \kappa_z^2}{\kappa_m^2}\right), \quad (9)$$

$$A(\mu) = \frac{\Gamma(\mu - 1)\cos(\mu\pi/2)}{4\pi^2}, \quad (10)$$

$$c(\mu) = \left[\frac{2\pi}{3}\Gamma\left(\frac{5-\mu}{2}\right)A(\mu)\right]^{\frac{1}{\mu-5}}, \quad (11)$$

$$\tilde{C}_n^2 = \beta C_n^2, \quad (12)$$

where, $\xi_x$ and $\xi_y$ denoting the anisotropic factors, $\kappa_0 = 2\pi/L_0$ with $L_0$ being the outer scale of atmospheric turbulence, and $\kappa_m = c(\alpha)/l_0$ with $l_0$ being the inner scale of atmospheric turbulence. $\mu$ is the generalized exponent parameter and $\Gamma(\cdot)$ is the gamma function. $\tilde{C}_n^2$ is the generalized refractive-index structure parameter with units $m^{3-\mu}$ used for indicating the intensity of atmospheric turbulence. $C_n^2$ is reduced to the structure parameter with unit [$m^{-2/3}$] and $\beta$ is a dimensional constant with unit [$m^{11/3-\mu}$] [26].

Substituting Eq. (3) and (8) into Eq. (6) and taking $m = 0, 1$ for the convenience of calculation, after the tedious integral, the analytical formulas for the elements of the CSDM for PCEShVB through the atmospheric turbulence can be obtained with the following expressions:

$$W_{\alpha\beta}(s_1, s_2; z) = A_\alpha A_\beta B_{\alpha\beta}\left(\frac{k}{2z}\right)^2 \exp\left[-\frac{ik}{2z}(s_1^2 - s_2^2)\right]$$
$$\times \exp\left[-\frac{1}{\rho_0^2}(s_1 - s_2)^2\right](W_{1\alpha\beta} + W_{2\alpha\beta} - W_{3\alpha\beta} - W_{4\alpha\beta}), \quad (13)$$

where

$$W_{j\alpha\beta} = \frac{1}{C_3^2}\exp\left(\frac{D_{2j}^2 + D_{4j}^2}{4B_2}\right)\exp\left[\frac{(AD_{4j} + 2BD_{3j})^2 + (AD_{2j} + 2BD_{1j})^2}{4BC}\right]$$
$$\times \left[\tau_{xx} + \tau_{yy} \pm i(\tau_{xy} - \tau_{yx})\right] \quad (j = 1 \sim 4), \quad (14)$$

$$\tau_{xx} = A + \frac{D_{4j}\left(AD_{4i} + 2BD_{3j}\right)}{2B} + \frac{A\left(AD_{4j} + 2BD_{3j}\right)^2}{2BC},$$

$$\tau_{yy} = A + \frac{D_{2j}\left(AD_{2j} + 2BD_{1j}\right)}{2B} + \frac{A\left(AD_{2j} + 2BD_{1j}\right)^2}{2BC},$$

$$\tau_{xy} = \frac{AD_{4j} + 2BD_{3j}}{2B}\left[D_{2j} + \frac{A\left(AD_{2j} + 2BD_{1j}\right)}{C}\right], \qquad (15)$$

$$\tau_{yx} = \frac{AD_{2j} + 2BD_{1j}}{2B}\left[D_{4j} + \frac{A\left(AD_{4j} + 2BD_{3j}\right)}{C}\right],$$

$$A = \frac{1}{\delta_{\alpha\beta}^2} + \frac{2}{\rho_0^2}, \quad B = \frac{1}{w_0^2} + \frac{ik}{2z} + \frac{A}{2}, \quad C = 4BB^* - A^2, \qquad (16)$$

$$D_{1j} = \frac{s_{1y} - s_{2y}}{\rho_0^2} - \frac{iks_{2y}}{z} + \Omega_{j,1}, \qquad D_{2j} = \frac{s_{2y} - s_{1y}}{\rho_0^2} + \frac{iks_{1y}}{z} + \Omega_{j,2},$$

$$D_{3j} = \frac{s_{1x} - s_{2x}}{\rho_0^2} - \frac{iks_{2x}}{z} + \Omega_{j,3}, \qquad D_{4j} = \frac{s_{2x} - s_{1x}}{\rho_0^2} + \frac{iks_{1x}}{z} + \Omega_{j,4}, \qquad (17)$$

$$\Omega = \begin{bmatrix} \Omega_0 & \Omega_0 & \Omega_0 & \Omega_0 \\ -\Omega_0 & -\Omega_0 & -\Omega_0 & -\Omega_0 \\ \Omega_0 & \Omega_0 & -\Omega_0 & -\Omega_0 \\ -\Omega_0 & -\Omega_0 & \Omega_0 & \Omega_0 \end{bmatrix}, \qquad (18)$$

$$T = \int_0^\infty \kappa^3 \Phi_n(\kappa) d\kappa, \qquad (19)$$

$$\frac{1}{\rho_0^2} = \frac{T\pi^2 k^2 z}{3}. \qquad (20)$$

Eq. (13)-(20) are the main propagation mathematical analytical formulas derived in this paper, which can be used to studied the statistical properties of the beam in atmospheric turbulence. And the turbulence quantity $T$ can be obtained by substituting Eq. (9)-(11) into Eq.(19) [26–28]:

$$T = \frac{\xi_x^2 + \xi_y^2}{2\xi_x^2 \xi_y^2} \frac{A(\mu)\tilde{C}_n^2}{2(\mu-2)} \left\{ \left[2\kappa_0^2 \kappa_m^{2-\mu} + (\alpha-2)\kappa_m^{4-\mu}\right] \exp\left(\frac{\kappa_0^2}{\kappa_m^2}\right) \Gamma\left(2 - \frac{\mu}{2}, \frac{\kappa_0^2}{\kappa_m^2}\right) - 2\kappa_0^{4-\mu} \right\}. \qquad (21)$$

where $\Gamma(\cdot, \cdot)$ is an incomplete gamma function.

The average intensity, degree of polarization (DOP) and the degree of coherence(DOC) of PCEShVB through the atmospheric turbulence can be expressed as [29–31]:

$$I(s;z) = W_{xx}(s;z) + W_{yy}(s;z), \qquad (22)$$

$$\Upsilon(s_1, s_2; z) = \frac{Tr\left[\overleftrightarrow{W}(s_1, s_2; z)\right]}{\sqrt{Tr\left[\overleftrightarrow{W}(s_1, s_1; z)\right] Tr\left[\overleftrightarrow{W}(s_2, s_2; z)\right]}}, \qquad (23)$$

$$P(s;z) = \sqrt{1 - \frac{4\text{Det}\left[\overleftrightarrow{W}(s;z)\right]}{\left\{Tr\left[\overleftrightarrow{W}(s;z)\right]\right\}^2}}, \qquad (24)$$

where Tr and Det represent the trace and the determinant of matrix, respectively. By substituting Eq. (13)-(20) into (22)-(24), accurate analytical expressions can be obtained. By studying the average intensity, coherence and polarization of PCEShVB on the receiving plane in free space and atmospheric turbulence, the necessary theoretical basis for the application of PCEShVB in free space communication and laser ranging is provided.

## 3. Numerical examples and discussions

In this section, numerical examples will be presented to analyze the propagation of the statistical characteristics of PCEShVB in atmospheric turbulence. To facilitate comparison, certain calculation parameters have been set as follows: $\mu=3.4$, $L_0=50$ m, $l_0=1$ mm, $\xi_x=2$, $\xi_y=3$, $\lambda=632.8$ nm, $w_0=10$ mm, $A_x=A_y=1$, $B_{xy}=0.25\exp(\pi/6)$. These values will be used in the calculation unless otherwise specified.

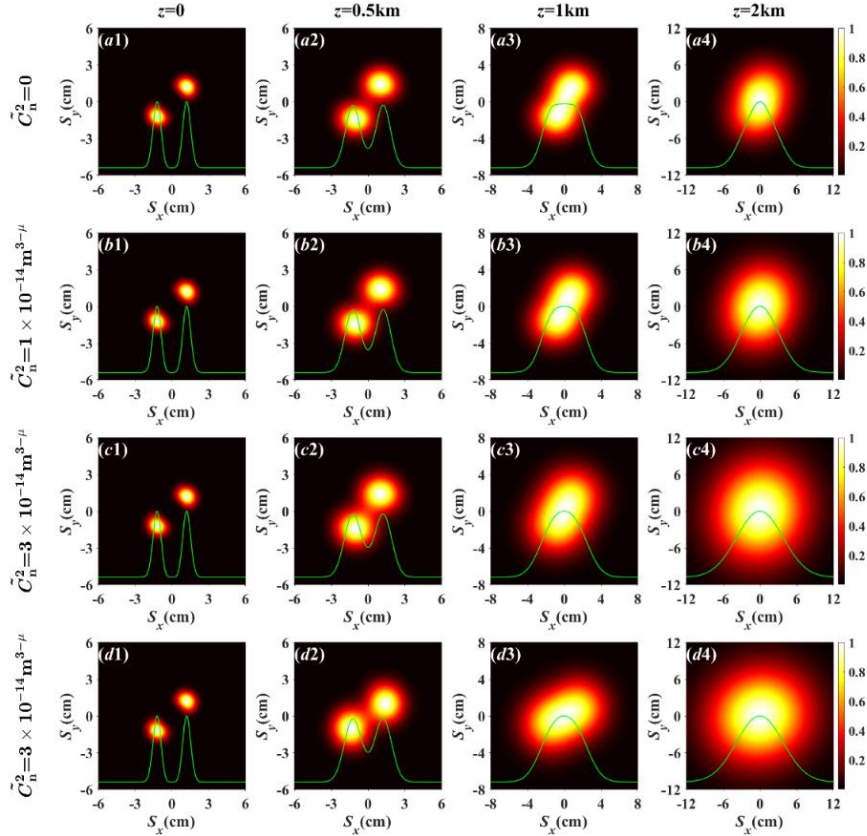

Fig. 1. The normalized average intensity distribution and diagonal line of PCEShVB through atmospheric turbulence with different $\tilde{C}_n^2$ at different propagation distance $z$; $\Omega_0=200$m$^{-1}$, $\delta_{xx}=\delta_{yy}=10$mm, (a)-(c) m=1, (d) m=-1.

Initially, an investigation is conducted on the normalized average intensity distribution of PCEShVB at various propagation distances through atmospheric turbulence with differing intensities, as illustrated in Fig. 1. The initial beam profile of PCEShVB is distinctly manifested as two spots along a 45° diagonal in Fig. 1 (a1) - (d1). And the two spots of the PCEShVB profile gradually begin to expand and merge with the increase of propagation distance, and eventually evolve into Gauss-like distribution. In addition, another interesting finding is that during the process of spot expansion and mergence, the overall distribution of the beam spot tends to rotate around the axis, rotating towards the *y*-axis when m=1, and towards the *x*-axis when m= −1. Other different vortex beams also have similar phenomena, which may be due to the fact that the vortex beam carries orbital angular momentum, and the direction of the vortex beam also affects its intensity distribution.

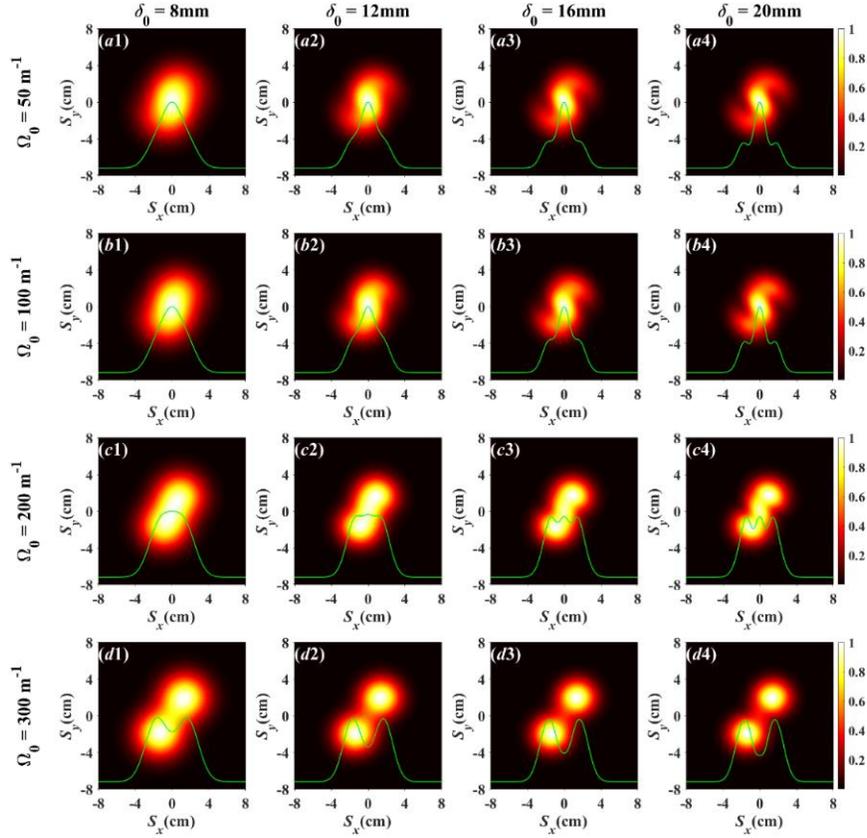

Fig. 2. The normalized average intensity distribution and diagonal line of PCEShVB with different $\delta_0$ and $\Omega_0$ through atmospheric turbulence at $z$ =1km; $\tilde{C}_n^2$=1×10$^{-15}$ m$^{3-\mu}$, m=1, $\delta_{xx}$=$\delta_{yy}$=$\delta_0$.

Fig.2 displays the normalized average intensity distribution and diagonal line of PCEShVB with different initial coherence length $\delta_0$ and hyperbolic sine parameter $\Omega_0$ through atmospheric turbulence at $z$ =1km. It can be found from Fig.2 (a1)- (a4) [or (b1)- (b4), (c1)- (c4), (d1)- (d4)] that under the same propagation distance, the degree of separation of the spots of PCEShVB is affected by the initial coherence length $\delta_0$, and the greater the initial coherence length, the greater the degree of separation of the spots. Moreover, From Fig.2 (a1)–(d1), it can be observed that as the hyperbolic sine parameter $\Omega_0$ gradually increases, the beam spot progressively expands, and eventually, focal spots separation occurs. Notably, at $\Omega_0$=50 m$^{-1}$, the overall distribution of PCEShVB exhibits orientation towards the y-axis, while at $\Omega_0$=300 m$^{-1}$, it remains relatively consistent with the initial plane. This observation suggests that an elevated $\Omega_0$ retards the evolution speed of PCEShVB. In summary, augmenting the initial coherence length $\delta_0$ and hyperbolic sine parameter $\Omega_0$ of PCEShVB can decelerate its evolution in atmospheric turbulence, conferring heightened resilience against atmospheric perturbations.

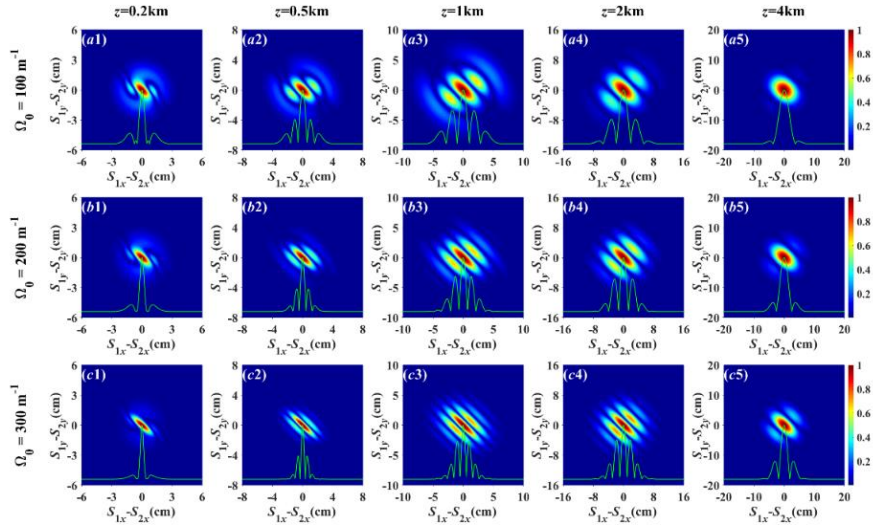

Fig. 3. The DOC distribution and diagonal line of PCEShVB with different $\Omega_0$ through atmospheric turbulence at different propagation distance $z$; $\tilde{C}_n^2=1\times10^{-15}$ m$^{3-\mu}$, $\xi_x=\xi_y=4$, m=1, $\delta_{xx}=\delta_{yy}$=10mm.

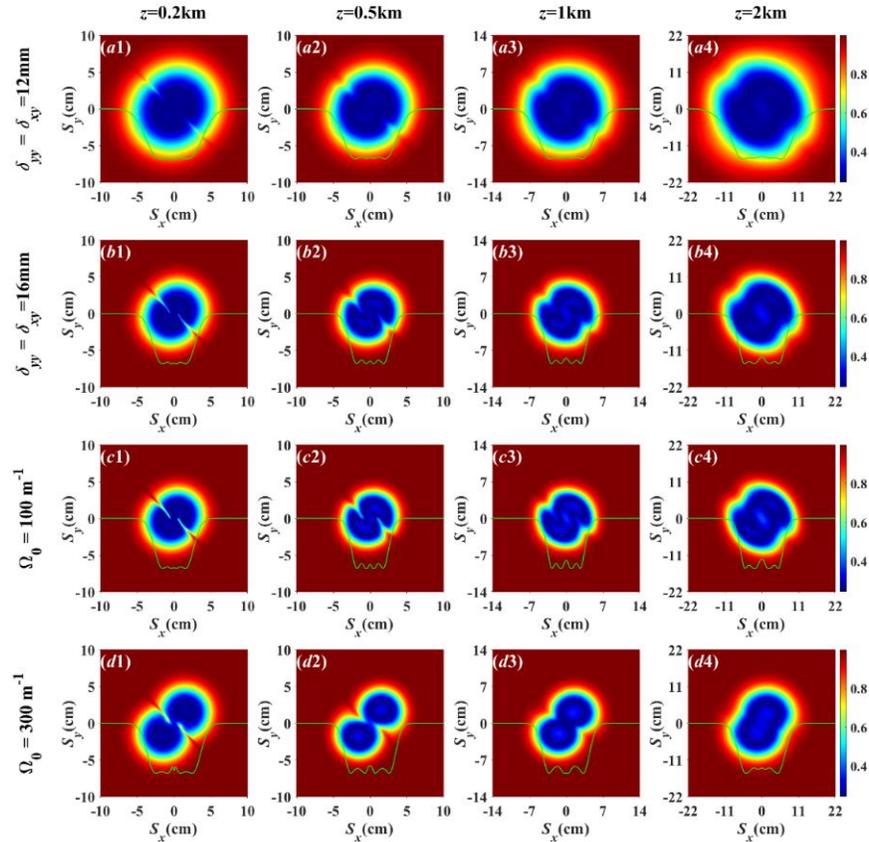

Fig. 4. The DOP distribution and diagonal line of PCEShVB with different $\Omega_0$ and different $\delta_{\alpha\beta}$ through atmospheric turbulence at different propagation distance $z$; $\tilde{C}_n^2=1\times10^{-15}$ m$^{3-\mu}$, m=1, $\delta_{xx}$=10mm, (c)-(d) $\delta_{yy}=\delta_{xy}$=20mm.

For the investigation of the coherence characteristics of the beam, the distribution of the DOC for PCEShVB with varying $\Omega_0$ in atmospheric turbulence has been depicted at different propagation distances in Fig. 3. It is well known that the initial DOC distribution of PCEShVB satisfies the Gaussian distribution. However, the result above shows that the DOC distribution of the beam begins to distort and develop a spiral distribution structure, as the propagation distance increases. With the further increase of the propagation distance, the DOC distribution appears similar to the interference fringe of light, and finally the fringe on both sides except the center begins to disappear, and the DOC evolves again into a Gaussian distribution (Gaussian-like distribution). In this interesting phenomenon, the quantity of interference fringes is contingent upon $\Omega_0$. A higher value of $\Omega_0$ correlates with an increased number of interference fringes, thereby necessitating a greater distance for the DOC of PCEShVB to transition into a Gaussian distribution.

Fig.4 displays the evolution of the DOP of PCEShVB with various initial coherent length $\delta_{\alpha\beta}$ and $\Omega_0$ in atmospheric turbulence. Primarily, the observation reveals that the evolution process of the DOP of PCEShVB closely mirrors the evolution of average intensity, as evidenced by the distribution of spiral drill bits and the occurrence of spot mergence. Then, it can be also found that when the propagation distance is short ($z$=0.2km), the overall DOP of PCEShVB is in a relatively small value. And the overall DOP of the beam will increase with the increase of the propagation distance. The DOP of PCEShVB with a larger initial coherence length is significantly larger [see Fig.4 (a1)- (a4), (b1)- (b4), (c1)- (c4)]. In addition, compared with Fig.4 (c1)- (c4) and (d1)- (d4), it can also be found that a large $\Omega_0$ can also lead to an increase in DOP of the beam. Furthermore, the center DOP of PCEShVB versus the propagation distance $z$ has been computed under various conditions, as illustrated in Fig. 5. It is shown that PCEShVB has a larger center DOP when propagating in the atmospheric turbulence with a small $\tilde{C}_n^2$ and large anisotropy factors $\xi_x$ ($\xi_y$). PCEShVB with a large initial coherent length will also give it a larger center DOP, and this is consistent with what is shown in Fig. 4. However, combined with Fig. 4(c1)- (c4),(d1)-(d4) and Fig. 5(c), it can be found that in short distance propagation, although PCEShVB with a large $\Omega_0$ will increase its overall DOP, it will result in a lower center DOP.

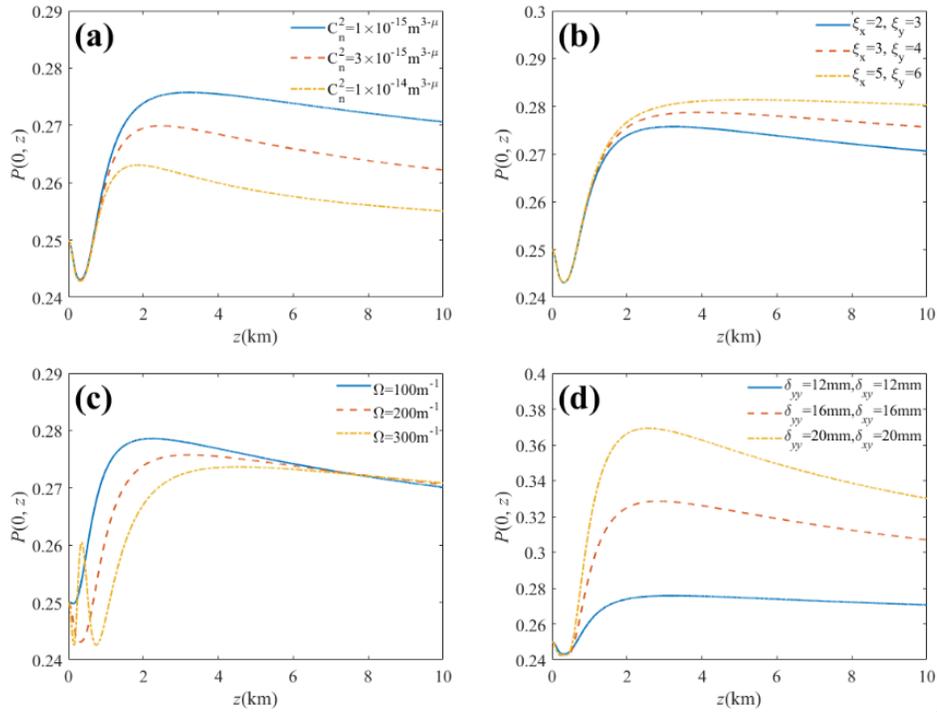

Fig. 5. The center DOP of PCEShVB vs propagation distance $z$ in different condition; m=1; (a) $\Omega_0$=200m$^{-1}$, $\delta_{xx}$=10mm, $\delta_{yy}$=$\delta_{xy}$=12mm; (b) $\Omega_0$=200m$^{-1}$, $\tilde{C}_n^2$=1×10$^{-15}$ m$^{3-\mu}$, $\delta_{xx}$=10mm, $\delta_{yy}$=$\delta_{xy}$=12mm; (c) $\tilde{C}_n^2$=1×10$^{-15}$ m$^{3-\mu}$, $\delta_{xx}$=10mm, $\delta_{yy}$=$\delta_{xy}$=12mm; (d) $\tilde{C}_n^2$=1×10$^{-15}$ m$^{3-\mu}$, $\Omega_0$=200m$^{-1}$.

## 4. Simulation

The random phase screen (RPS) is a well-established method for simulating the propagation of a beam in atmospheric turbulence. Consequently, the RPS technique is employed to simulate the propagation characteristics of PCEShVB in atmospheric turbulence. The fundamental concept underlying RPS involves treating the beam propagation process in atmospheric turbulence as two distinct yet simultaneous phenomena: vacuum propagation and phase modulation within the atmospheric turbulence medium. This approach allows us to substitute the continuous random medium along the propagation path with a series of phase screens (PS) spaced at intervals of $\Delta z$, as depicted in Fig. 6. The beam on the source plane passes through these PSs in turn to simulate the turbulence modulation effect, and finally reaches the receiving surface.

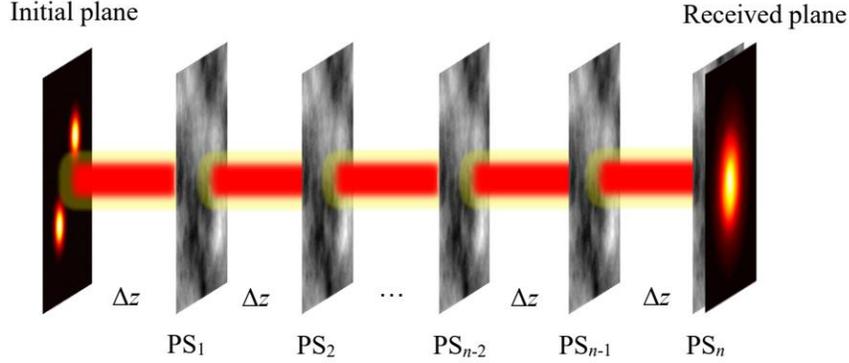

Fig.6 Schematic diagram of multi-phase screen method. PS: phase screen.

The simulation process can be characterized by [22]:

$$E_i(x, y; z_i) = F^{-1}\left\{F\left\{E_{i-1}(x, y; z_{i-1})\exp[i\phi(x, y; z_i)]\right\}\exp\left[-\frac{i\Delta z}{2k}\left(\kappa_x^2 + \kappa_y^2\right)\right]\right\}, \quad (25)$$

where, $z_{i-1}$ and $z_i$ represent the position of the ($i$–1)-th and $i$-th PS, $E_{i-1}(x, y; z_{i-1})$ and $E_i(x, y; z_i)$ is the fully coherent light field on the ($i$–1)-th and $i$-th PS, $\phi(x, y; z_i)$ denotes phase distribution on the $i$-th PS, $F$ and $F^{-1}$ denote the Fourier transform and inverse Fourier transform, respectively. There are two key points in simulation: (1) how to synthesize turbulent phase with turbulent information; (2) how to extend RPS to the partially coherent beam field [30,32].

The current commonly used method for constructing a PS is to use the turbulent refractive index spectrum and a complex Gaussian random matrix to generate a complex random field in phase space, and then perform inverse Fourier transform to obtain the spatial distribution of the two-dimensional phase. This method is called the power spectrum inversion method, also known as the Fourier transform method. The process of PS generation is as follows.

The relationship between the phase power spectrum $\Phi_\theta(\kappa_x, \kappa_y)$ caused by turbulence and the turbulent refractive index spectrum $\Phi_n(\kappa_x, \kappa_y)$ is:

$$\Phi_\theta\left(\kappa_x, \kappa_y\right) = 2\pi k^2 \Delta z \Phi_n\left(\kappa_x, \kappa_y\right), \quad (26)$$

A two-dimensional frequency domain complex random phase field $\phi'(\kappa_x, \kappa_y)$ in phase space can be obtained by filtering the phase power spectrum with a set of complex Gaussian random numbers $a_r$:

$$\phi'\left(\kappa_x, \kappa_y\right) = a_r\sqrt{\Phi_\theta\left(\kappa_x, \kappa_y\right)}, \quad (27)$$

$$a_r = A_r + iB_r, \quad (28)$$

where $A_r$ and $B_r$ are Gaussian random numbers with mean and variance both equal to zero and one, respectively. By performing the inverse Fourier transform on $\phi'(\kappa_x, \kappa_y)$, a two-dimensional spatial complex random phase field $\phi(x, y)$ can be obtained:

$$\phi(x,y) = \iint a_r \sqrt{\Phi_\theta(\kappa_x,\kappa_y)} \exp\left[i(\kappa_x x + \kappa_y y)\right] d\kappa_x d\kappa_y. \tag{29}$$

The simulation of PC beam is generally carried out using the complex screen method. The complex screen method involves placing a complex screen (CS) with certain statistical characteristics on the source plane of the fully coherent beam to synthesize the PC beam. The generation of the CS is similar to that of a PS, but it changes its spectral density function. The CS can modulate both the amplitude and phase of the light source simultaneously, and its correlation function is strictly equal to that of the PC beam. Therefore, the CS can be used to simulate PC beams with arbitrary coherence structures. The principle of its generation is as follows.

A fully coherent beam modulated by a CS can be expressed as:

$$E(x,y) = E_0(x,y) F(x,y). \tag{30}$$

where $E_0(x, y)$ is the incident electric field of a completely coherent beam, $F(x, y)$ denotes the CS with certain spatial correlation characteristics.

First, the coherence function $g(\Delta r_x, \Delta r_y)$ is Fourier transformed to obtain its spatial spectrum:

$$S(f_x, f_y) = \iint g(\Delta r_x, \Delta r_y) \exp\left[2\pi i(f_x \Delta r_x + f_y \Delta r_y)\right] d\Delta r_x d\Delta r_y, \tag{31}$$

where $\Delta r_x = x_1-x_2$, $\Delta r_y = y_1-y_2$, $f_x$ and $f_y$ are spatial frequencies. In the Fourier domain, we can use the product of the square root of the power spectral density function and the complex Gaussian random number of zero mean and unit variance to represent the CS, so we also use the complex Gaussian random number $a_r$ to filter $S(f_x, f_y)$:

$$F'(f_x, f_y) = a_r(f_x, f_y) \sqrt{S(f_x, f_y)}. \tag{32}$$

Subsequently, the computation of the CS $F(x, y)$ is attainable through the inverse Fourier transformation of the modified spectrum $F'(f_x, f_x)$. And this method can be also extended to generate vector PC beam just by modulating each component of electric field of the vector beam with CS, respectively.

In summary, the logical process of numerical simulation of PC beam in atmospheric turbulence can be summarized as follows:

(1) Generate a set of PSs and fix them on the propagation path.

(2) Generate another CS and fix it on the source plane of fully coherent beam. The electric field passes through the CS and then sequentially passes through the PSs fixed on the propagation path to reach the receiving plane.

(3) Repeat step (2) $K_1$ times to obtain $K_1$ frames of electric field passing through the CS and PSs. Average these $K_1$ frames of electric field to obtain one frame of PC beams passing through atmospheric turbulence.

(4) Repeat steps (1)-(3) $K_2$ times to obtain $K_2$ frames of PC beams passing through atmospheric turbulence. These $K_2$ frames can be used to analyze and calculate the statistical characteristics of PC beams after passing through atmospheric turbulence, including intensity, coherence, polarization, and other statistical properties.

In the simulation, the distance between each two PS is set to $\Delta z$=50m, and the number of data acquisition samples is set to 512×512. And we take $K_1$=$K_2$=200, which means that both PSs and CS have changed 200 times.

As illustrated in Fig. 7 and Fig. 8, the simulation results for light intensity and DOC are presented, employing identical calculation parameters as in Fig. 1 and Fig. 3. The simulation results obtained via the RPS method demonstrate significant agreement with the theoretical predictions, thereby confirming the precision of the theoretical findings.

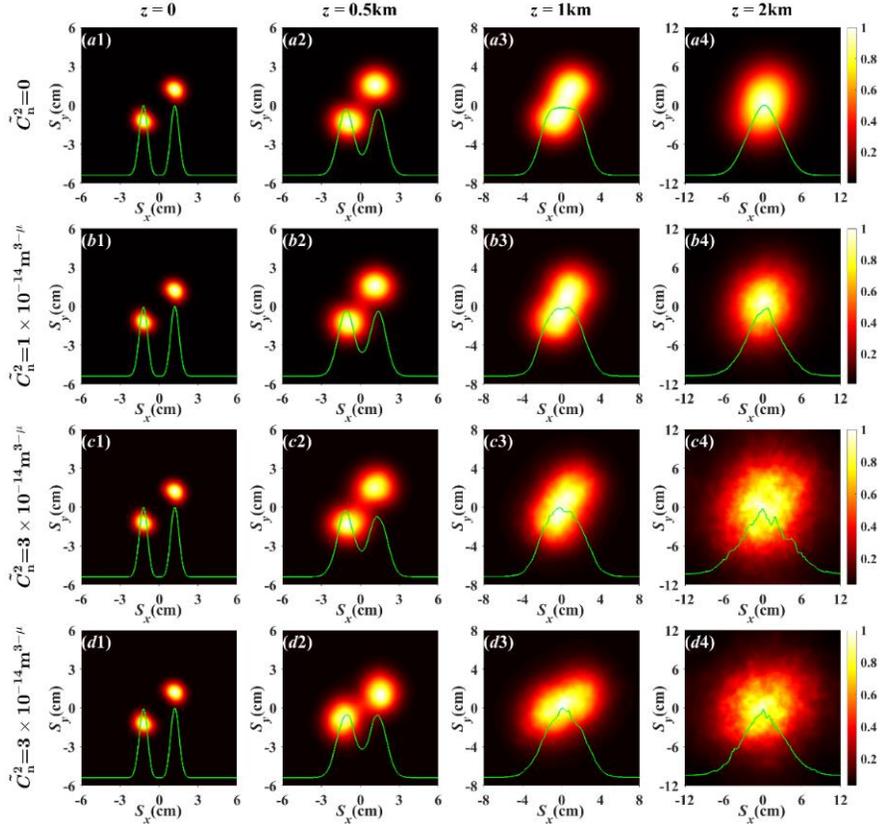

Fig.7 The simulation of Fig.1

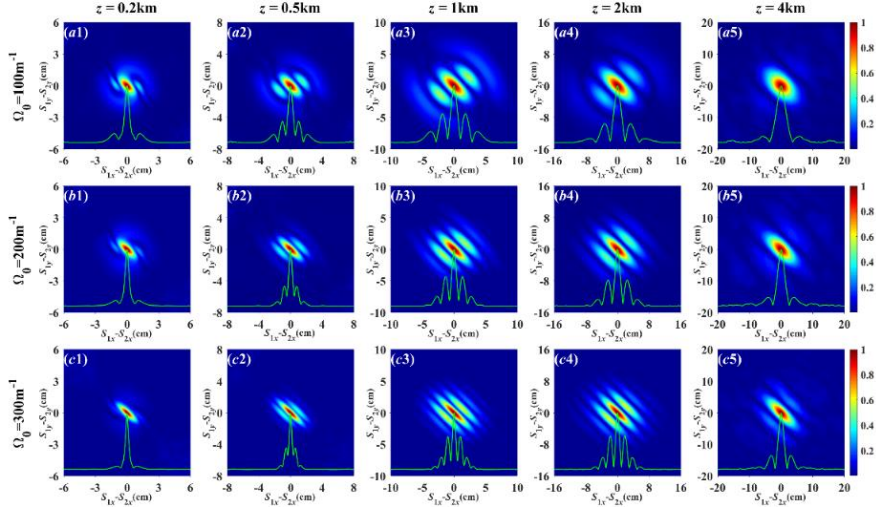

Fig.8 The simulation of Fig.3

## 5. Conclusions

In conclusion, this paper presents the derived analytical propagation expressions for the PCEShVB through anisotropic atmospheric turbulence and investigates the evolution characteristics of propagation, including average intensity, the DOC and the DOP. Notably, intriguing phenomena have been observed wherein the overall distribution of light spots exhibits a tendency to rotate around the axis during propagation. When m=1, the light spots exhibit rotation towards the *y*-axis, and when m= −1, the rotation occurs towards the *x*-axis. Additionally, PCEShVB with larger initial coherence length $\delta_0$ and $\Omega_0$ experiences slower evolution in atmospheric turbulence and demonstrates greater resistance to atmospheric turbulence. Furthermore, the distribution of DOC of PCEShVB appears similar to the pattern of interference fringe of light in propagating. Moreover, the number of 'interference fringes' increases with the increase of hyperbolic sine parameter $\Omega_0$. It has also been identified that PCEShVB with larger initial coherence length $\delta_{\alpha\beta}$ and hyperbolic sine parameter $\Omega_0$ results in increased spot separation and higher overall DOP. In addition, the random phase screen method was applied to simulate the propagation of PCEShVB in atmospheric turbulence. The obtained simulation results exhibit consistent concordance with the theoretical findings, thereby substantiating the validity of the proposed theoretical framework.

**Funding**

National Natural Science Foundation of China (No. 12003065&12033009);
Major Science and Technology projects in Yunnan Province (No.2019FA002);

**Disclosures**

The authors declare no conflicts of interest.

**Data availability**

No data were generated or analyzed in the presented research.

**References**

1. V. Garcés-Chávez, D. McGloin, H. Melville, W. Sibbett, and K. Dholakia, "Simultaneous micromanipulation in multiple planes using a self-reconstructing light beam," Nature **419**, 145–147 (2002).
2. Z. Qiao, Z. Wan, G. Xie, J. Wang, L. Qian, and D. Fan, "Multi-vortex laser enabling spatial and temporal encoding," PhotoniX **1**, 1–14 (2020).
3. H. Wu, Y. Dan, X. Yu, Y. Xu, and N. Deng, "Effect of outer scale on partially coherent Sinh-Gaussian beams in inhomogeneous atmospheric turbulence," J. Mod. Opt. **67**, 890–898 (2020).
4. Y. Dan, Y. Ai, and Y. Xu, "Propagation properties and turbulence distance of partially coherent Sinh-Gaussian beams in non-Kolmogorov atmospheric turbulence," Optik **127**, 9320–9327 (2016).
5. H. Ma, J. Li, and P. Sun, "Effective tensor approach for simulating the propagation of partially coherent Hermite–sinh–Gaussian beams through an ABCD optical system in turbulent atmosphere," JOSA A **36**, 2011–2016 (2019).
6. M. Bayraktar, "Propagation of chirped sinh-Gaussian beams in uniaxial crystals orthogonal to the optical axis," Opt. Quantum Electron. **53**, 512 (2021).
7. K. Huang, Y. Xu, J. Cao, and Y. Li, "Propagation Characteristics of Partially-Coherent Radially-Polarized Vortex Beams Through Non-Kolmogorov Turbulence Along a Slant Path," J. Russ. Laser Res. **44**, 110–120 (2023).
8. Y. Cai and S. Zhu, "Orbital angular moment of a partially coherent beam propagating through an astigmatic ABCD optical system with loss or gain," Opt. Lett. **39**, 1968–1971 (2014).
9. Y. Bai, H. Lv, X. Fu, and Y. Yang, "Vortex beam: generation and detection of orbital angular momentum [Invited]," Chin. Opt. Lett. **20**, 012601 (2022).
10. W. Cheng, J. W. Haus, and Q. Zhan, "Propagation of vector vortex beams through a turbulent atmosphere," Opt. Express **17**, 17829–17836 (2009).
11. L. W. Casperson, D. G. Hall, and A. A. Tovar, "Sinusoidal-Gaussian beams in complex optical systems," JOSA A **14**, 3341–3348 (1997).
12. Y. Huang, F. Wang, Z. Gao, and B. Zhang, "Propagation properties of partially coherent electromagnetic hyperbolic-sine-Gaussian vortex beams through non-Kolmogorov turbulence," Opt. Express **23**, 1088 (2015).
13. H. T. Eyyuboğlu and Y. Baykal, "Hermite-sine-Gaussian and Hermite-sinh-Gaussian laser beams in turbulent atmosphere," JOSA A **22**, 2709–2718 (2005).
14. Y. Zhang, X. Zhou, and X. Yuan, "Performance analysis of sinh-Gaussian vortex beams propagation in turbulent atmosphere," Opt. Commun. **440**, 100–105 (2019).
15. M. Bayraktar, "Average intensity of astigmatic hyperbolic sinusoidal Gaussian beam propagating in oceanic turbulence," Phys. Scr. **96**, 025501 (2021).

16. M. Bayraktar, "Propagation of partially coherent hyperbolic sinusoidal Gaussian beam in biological tissue," Optik **245**, 167741 (2021).
17. L. Biferale and I. Procaccia, "Anisotropy in turbulent flows and in turbulent transport," Phys. Rep. **414**, 43–164 (2005).
18. I. Toselli, L. C. Andrews, R. L. Phillips, and V. Ferrero, "Angle of arrival fluctuations for free space laser beam propagation through non kolmogorov turbulence," in *Atmospheric Propagation IV* (SPIE, 2007), Vol. 6551, pp. 149–160.
19. I. Toselli, "Introducing the concept of anisotropy at different scales for modeling optical turbulence," JOSA A **31**, 1868–1875 (2014).
20. F. Flossmann, U. T. Schwarz, and M. Maier, "Propagation dynamics of optical vortices in Laguerre–Gaussian beams," Opt. Commun. **250**, 218–230 (2005).
21. E. Wolf, "Unified theory of coherence and polarization of random electromagnetic beams," Phys. Lett. A **312**, 263–267 (2003).
22. K. Yong, S. Tang, X. Yang, and R. Zhang, "Propagation characteristics of a ring Airy vortex beam in slant atmospheric turbulence," JOSA B **38**, 1510–1517 (2021).
23. L. Zhao, Y. Xu, N. Yang, Y. Xu, and Y. Dan, "Propagation factor of partially coherent radially polarized vortex beams in anisotropic turbulent atmosphere," JOSA A **38**, 1255–1263 (2021).
24. Y. Xu, Y. Li, and X. Zhao, "Intensity and effective beam width of partially coherent Laguerre–Gaussian beams through a turbulent atmosphere," JOSA A **32**, 1623–1630 (2015).
25. K. Huang, Y. Xu, L. Zhao, J. Cao, and Y. Li, "Propagation factor of partially coherent vector vortex beam in inhomogeneous turbulent atmosphere," Optik **271**, 170247 (2022).
26. J. Wang, S. Zhu, H. Wang, Y. Cai, and Z. Li, "Second-order statistics of a radially polarized cosine-Gaussian correlated Schell-model beam in anisotropic turbulence," Opt. Express **24**, 11626 (2016).
27. Y. Ata, Y. Baykal, and M. C. Gökçe, "Average channel capacity in anisotropic atmospheric non-Kolmogorov turbulent medium," Opt. Commun. **451**, 129–135 (2019).
28. M. Cheng, L. Guo, J. Li, and Q. Huang, "Propagation properties of an optical vortex carried by a Bessel–Gaussian beam in anisotropic turbulence," JOSA A **33**, 1442–1450 (2016).
29. J. Li and B. Lü, "Propagation of Gaussian Schell-model vortex beams through atmospheric turbulence and evolution of coherent vortices," J. Opt. Pure Appl. Opt. **11**, 045710 (2009).
30. L. Zhao, Y. Xu, and Y. Dan, "Evolution properties of partially coherent radially polarized Laguerre-Gaussian vortex beams in an anisotropic turbulent atmosphere," Opt. Express **29**, 34986–35002 (2021).
31. L. Zhao, Y. Xu, and S. Yang, "Statistical properties of partially coherent vector beams propagating through anisotropic atmospheric turbulence," Optik **227**, 166115 (2021).
32. J. Yu, Y. Huang, F. Wang, X. Liu, G. Gbur, and Y. Cai, "Scintillation properties of a partially coherent vector beam with vortex phase in turbulent atmosphere," Opt. Express **27**, 26676–26688 (2019).